\documentclass[aps,prb,groupedaddress,twocolumn,showpacs]{revtex4}
\usepackage{graphicx,psfrag,amsmath,calc,amssymb}

\begin{document}

\title{The Lifshitz Transition in $d$-wave Superconductors}
\author{S. S. Botelho}
\author{C. A. R. S\'{a} de Melo}
\affiliation{School of Physics, Georgia Institute of Technology,
             Atlanta Georgia 30332}

\date{\today}

\begin{abstract}
The BCS to BEC evolution has been recently the focus of studies in 
superconductors and cold atomic gases. For a $d$-wave system, we show that 
a Lifshitz transition occurs at a critical particle density which separates
two topologically distinct phases according to their quasiparticle 
excitation energies: a BCS-like gapless superconductor in the {\it higher} 
density limit and a BEC-like fully gapped superconductor in the {\it lower}
density limit. This transition is second order according to Ehrenfest's 
classification, but it occurs without a change in the symmetry of the 
order parameter, and thus can not be classified under Landau's scheme. 
To illustrate the nature of the transition, we compute the compressibility 
and the superfluid density as functions of particle density.
\end{abstract}

\maketitle


\section{Introduction}

The evolution from BCS to Bose-Einstein condensation (BEC) 
superconductivity/superfluidity has attracted considerable amount of 
interest in the condensed 
matter~\cite{leggett-80a,leggett-80b,nozieres-85,sademelo-93,zwerger-97}
and atomic physics~\cite{holland-01,griffin-02,jin-04} communities. 
In the atomic physics community, this interest resulted from 
the possibility of studying condensation phenomena 
in fermionic atomic gases,
where the scattering length (and, thus, the effective interaction strength)
can be tunned via Feshbach resonances for a 
given density~\cite{holland-01,griffin-02,jin-04}.  
In the condensed matter physics community, the interest in the 
BCS to BEC evolution arouse in the context of high-$T_c$
superconductivity~\cite{sademelo-93, zwerger-97}, where the nature of 
the superconducting and normal states
changes as function of carrier concentration. 
Furthermore, experimentalists seem
to start having some control over the carrier concentration 
using electronic doping via ferroelectric oxides~\cite{ahn-03}. 
The fundamental issue that needs to be addressed is whether there is a 
quantum phase transition in the evolution from a BCS to BEC ground state, 
as particle density or scattering length (interaction strength) are varied. 
In this paper we show that for a two-dimensional $d$-wave superconductor
there is a Lifshitz transition between the BCS and BEC 
ground states. 
The transition is second order according to Ehrenfest's classification of 
phase transitions, but it occurs without change in symmetry as Landau's 
classification would demand. The quantum phase transition found 
has a logarithmic singularity in the 
compressibility at a critical concentration for fixed interaction.
Furthermore, near the critical point the low temperature 
superfluid density exhibits a dramatic change in behavior. 

This manuscript is organized as follows. In Section II, we discuss the
Hamiltonian and the interparticle potential used in our model. 
In Section III, our functional integral calculation is presented, 
and the order parameter and number equations are derived at the 
mean-field level. The analysis of Gaussian fluctuations about the
saddle point solution is also performed in this section. 
In Section IV, our results for the electronic compressibility and
the phase diagram of the system are shown, while in Section V an
analogy with the Lifshitz transition in metals is discussed.
The superfluid density is analyzed in Section VI and, finally, our 
concluding remarks are summarized in Section VII.


\section{Hamiltonian and Interaction Potential}

We study a two-dimensional continuum model of fermions 
of mass $m$ described by the Hamiltonian 
($\hslash=k_B=1$)
\begin{equation}
{\cal H}=\sum_{{\bf k},\sigma}\epsilon_{\bf k}\psi_{{\bf k}\sigma}^\dagger
\psi_{{\bf k}\sigma} + \sum_{{\bf k},{\bf k}',{\bf q}} V_{{\bf k}{\bf k}'}
b_{{\bf k}{\bf q}}^\dagger b_{{\bf k}'{\bf q}} ,
\end{equation}
where $b_{{\bf k}{\bf q}}=\psi_{-{\bf k}+{\bf q}/2\uparrow}
\psi_{{\bf k}+{\bf q}/2\downarrow}$
and
$\epsilon_{\bf k}=k^2/2m$.
We consider the following  separable potential in ${\bf k}$-space, 
\begin{equation}
V_{{\bf k}{\bf k}'} = -\lambda_d \Gamma ({\bf k})
\Gamma ({\bf k}'),
\end{equation}
which includes only the dominant angular momentum channel, 
assumed to be $d$-wave.
The interaction term can be written as
$\Gamma ({\bf k}) = h (k) g(\hat{{\bf k}})$, where 
$h (k)=(k/k_1)^2/[1+k/k_0]^{5/2}$ controls the range 
of the interaction, $g (\hat{{\bf k}})=\cos(2\varphi)$ sets its angular 
dependence, and $\varphi$ is the momentum angle in 
polar coordinates~\cite{randeria-90}.
In this case, $k_0\sim R_0^{-1}$, where $R_0$ plays the role 
of the interaction range, and both $k_0$ and $k_1$ set the momentum scales 
in the short and long wavelentgh limits.
We work under the assumption that the system is dilute 
enough, i.e., $k_{F_{\rm max}}^2\ll k_0^2$.
When computing physical properties throughout the manuscript, 
we scale momenta by $k_{F_{\rm max}}$, energies by 
$\epsilon_{F_{\rm \max}} = k^2_{F_{\rm max}}/2m$, 
velocities by $v_{F_{\rm max}} = k_{F_{\rm max}}/m$,
and particle density $n$ by $n_{\rm max}/2\pi$,
where $n_{\rm max} = k^2_{F_{ \rm max}}/2\pi$. 
If we choose, for instance, $k_0=\sqrt{10}{\rm \AA^{-1}}$ 
($R_0\approx 0.32{\rm \AA}$) 
and define $k_{F_\mathrm{max}}=k_0/10$, then 
$n_{\mathrm{max}}\approx 1.59\times 10^{14}\,\mathrm{cm}^{-2}$. 
We next discuss the effective action and analyze the effects
of Gaussian fluctuations about the saddle point solution.


\section{Effective Action and Gaussian Fluctuations}
The partition function $Z$ at a temperature $T = \beta^{-1}$ is written as an 
imaginary-time functional integral with action
$
S=\int_0^\beta d\tau[\sum_{{\bf k},\sigma} \psi_{{\bf k}\sigma}^\dagger(\tau)
(\partial_\tau -\mu)\psi_{{\bf k}\sigma}(\tau) + {\cal H}].
$
Introducing the usual Hubbard-Stratonovich field $\phi_{\bf q}(\tau)$, 
which couples to $\psi^\dagger\psi^\dagger$, and integrating out the fermionic 
degrees of freedom, we obtain 
$
Z=\int {\cal D}\phi{\cal D}\phi^* \,
\exp(-S_\mathrm{eff}[\phi,\phi^*]),
$
with the effective action given by
$$
S_{\mathrm{eff}} = \int_0^\beta \! d\tau
\left[
U (\tau) +
\sum_{{\bf k},{\bf k'}} \left(\xi_{\bf k} \delta_{{\bf k},{\bf k'}} -
{\rm Tr} \ln\mathbf{G}_{{\bf k},{\bf k'}}^{-1}(\tau)\right)
\right], 
$$
where $\xi_{\bf k} = \epsilon_{\bf k} - \mu$, 
$
U (\tau) =
\sum_{\bf k} | \phi_{\bf k}(\tau)|^2 / \lambda ,
$
and $\mathbf{G}_{{\bf k},{\bf k'}}^{-1}(\tau)$ is the (inverse) Nambu propagator,
\begin{equation}
\mathbf{G}_{{\bf k},{\bf k'}}^{-1}(\tau) = 
\left(
\begin{array}{cc}
-(\partial_\tau + \xi_{\bf k}) \delta_{{\bf k},{\bf k'}}  & 
\Lambda_{{\bf k},{\bf k'}}(\tau) \\[3mm]
\Lambda^*_{{\bf k'},{\bf k}}(\tau) & 
-(\partial_\tau - \xi_{\bf k}) \delta_{{\bf k},{\bf k'}}
\end{array}
\right),
\end{equation}
with 
$
\Lambda_{{\bf k},{\bf k'}}(\tau) = 
\phi_{\bf k-k'}(\tau) \Gamma(({\bf k+k'})/2) .
$

To study the BCS to BEC evolution it is necessary to go beyond the
standard saddle point approximation, and include at
least Gaussian fluctuations~\cite{sademelo-93,zwerger-97}. Assuming  
$
\phi_{\bf q}(\tau) = 
\Delta_0 \delta_{{\bf q},0} + \eta_{\bf q} (\tau) ,
$
and performing an expansion in $S_{\rm eff}$ to quadratic order
in $\eta$, one obtains
\begin{equation}
S_{\rm {Gauss}} = S_{0}[\Delta_0] + 
{1\over 2} \sum_q \underline\eta^\dagger(q) {\bf M}(q) 
\underline\eta(q) ,
\end{equation}
where $S_{0}$ is the saddle point action, 
$\underline\eta^\dagger = [\eta^*(q), \eta(-q)]$, and 
$q \equiv ({\bf q}, iq_m)$ with $iq_m=i2m\pi/\beta $. 
The inverse fluctuation propagator ${\bf M}$ is a $2\times 2$ 
matrix to be defined later. 

\begin{figure*}
\begin{center}
\includegraphics[width=4cm]{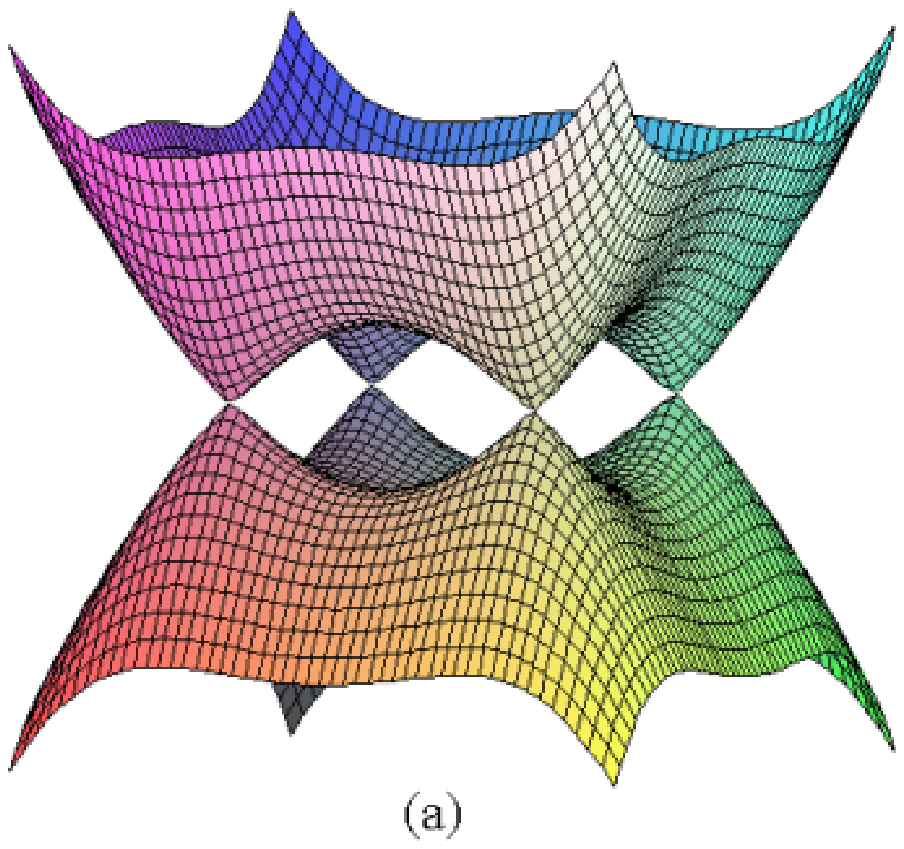}
\hspace{5mm}
\includegraphics[width=4cm]{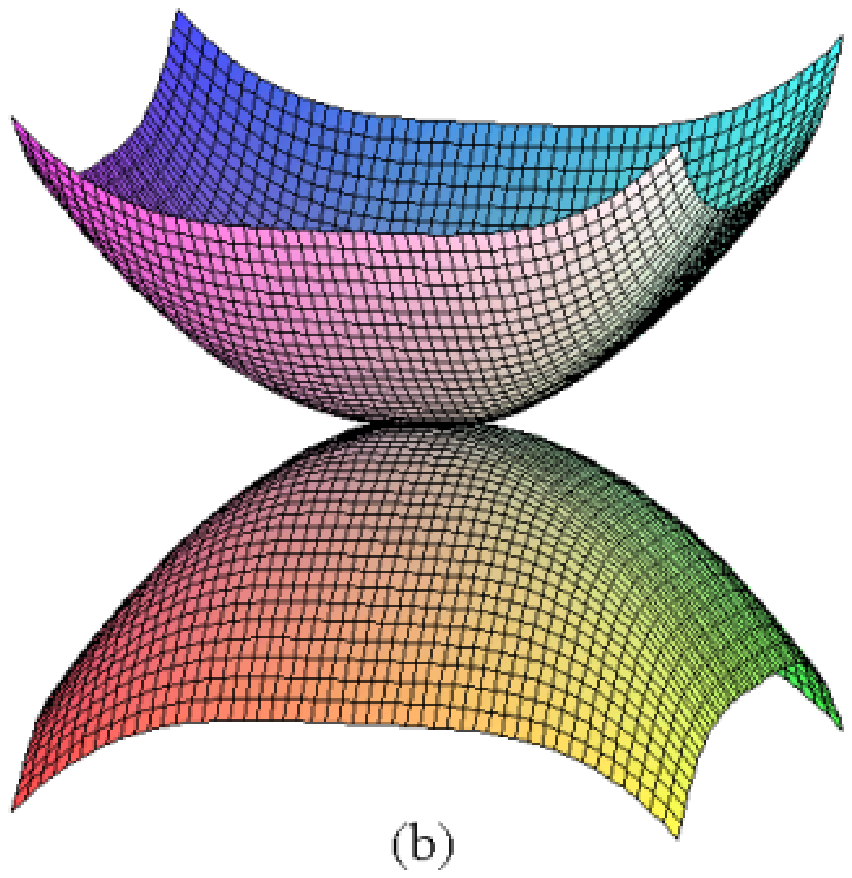}
\hspace{5mm}
\includegraphics[width=4cm]{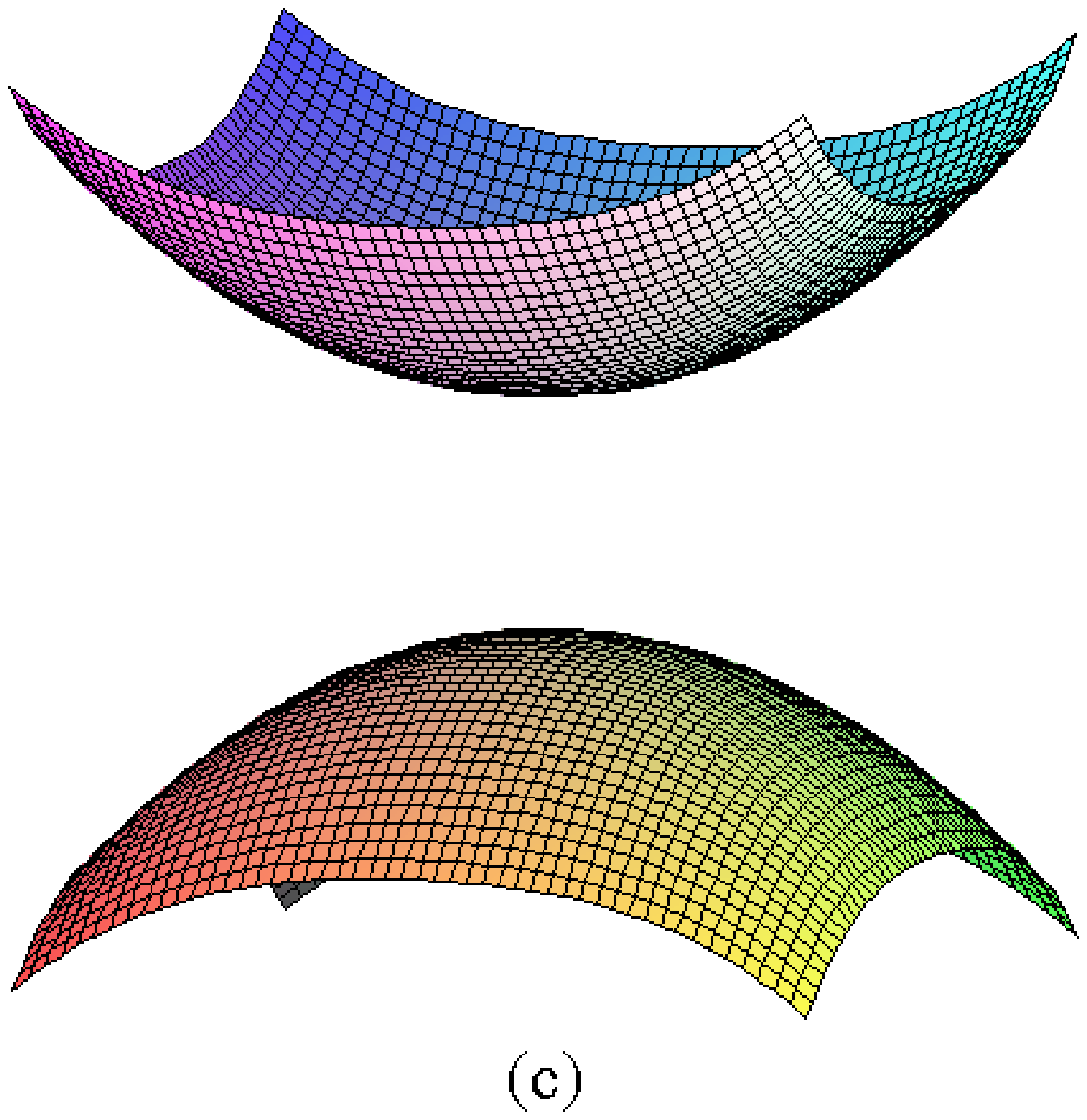}
\end{center}
\vspace{-5mm}
\caption{\small Plot of $\pm E({\bf k}, \mu)$ for (a) $\mu>0$, 
(b) $\mu=0$ and (c) $\mu<0$. Notice the collapse of the 
Dirac cones at $\mu=0$.}
\label{fig:excitation-energy}
\end{figure*}

{\it The Saddle Point Equation:} Starting with the 
stationarity condition 
$[\delta S_{\rm eff}/
\delta\phi_{\bf q}^*(\tau')]_{\Delta_0} = 0$,  
Fourier transforming from imaginary time to Matsubara frequency, 
$ik_n = i(2n+1)\pi/\beta$, and performing the frequency sum, the saddle point 
condition can be written as
\begin{equation}
{1\over\lambda_d} = \sum_{\bf k} {\Gamma^2({\bf k}) \over 2E_{\bf k} } 
\tanh\left(\beta E_{\bf k}\over 2\right),
\end{equation}
where 
$E_{\bf k} = \sqrt{\xi_{\bf k}^2+ \Delta_0^2 \Gamma^2({\bf k})}$
is the quasiparticle excitation energy, with 
$\xi_{\bf k}=\epsilon_{\bf k} - \mu$.
Fig.~\ref{fig:excitation-energy} shows a plot of $\pm E({\bf k}, \mu)$ 
as a function of ${\bf k}$ for three different values of the chemical
potential $\mu$. Notice that the Dirac cones collapse at $\mu = 0$.
%

{\it The Number Equation:} Using the thermodynamic relation 
$N =-\partial\Omega/\partial\mu$, and taking
$\Omega_{\rm Gauss}  = \Omega_0  + \Omega_{\rm fluct}$,  
where 
$\Omega_0 = S_{0}[\Delta_0] /\beta$ 
and
$\Omega_{\rm fluct} = \beta^{-1} \sum_q \ln {\rm det}[{\bf M} (q)]$, 
one can write the number equation as
\begin{equation}
\label{eqn:number-equation}
N_{\rm Gauss} = N_0 + N_{\rm fluct},
\end{equation}
where
$
N_0 = - \partial \Omega_0 / \partial \mu = 2 \sum_{\bf k} n_{\bf k},
$ 
with
$
n_{\bf k} = 1/2 \left[ 1  - {\xi_{\bf k} 
\tanh(\beta E_{\bf k}/2) / E_{\bf k}}
\right]
$
being the momentum distribution, and
\begin{equation}
N_{\rm fluct} = - { \partial \Omega_{\rm fluct} \over  \partial \mu } =
T \sum_{\bf q}\sum_{i q_n}
\left[ { 
{-\partial ({\rm det} {\bf M})/ \partial \mu } \over 
{\rm det} {\bf M} ({\bf q}, iq_n)
}
\right]
\end{equation}
being the fluctuation contribution to $N_{\rm Gauss}$. 
At $T=0$, we find well defined Goldstone modes for all couplings,
provided that $|{\bf q}|$ is sufficiently small. This collective
mode appears as a pole in the two-particle excitation 
spectrum determined by ${\rm det} [ {\bf M} ({\bf k}, z)] = 0$,
with $iq_n \to z$.
This pole has the form $z = c |{\bf q}| - i d |{\bf q}|^2$, 
where $c>0$ is the speed of sound and $d \ge 0$ is the damping coefficient.
For $\mu < 0$, $d$ vanishes and the contribution from the collective 
mode pole $z = c|{\bf q}|$ dominates at sufficiently low temperatures, 
since the two-particle excitations are gapped. 
For $\mu > 0$, $d$ becomes positive, and the spectrum of two-particle 
excitations is gapless due to the presence of the Dirac points. 
Thus, unlike the $s$-wave case, Landau damping occurs even at $T = 0$. 
However, for small $|{\bf q}|$, the Goldstone mode is underdamped, i.e.,
the real part of the pole dominates. 
Therefore, for either $\mu>0$ or $\mu<0$, these fluctuation effects lead to 
\begin{equation}
{N_{\rm fluct} \over N_{\rm max} } \sim  
- {\zeta(3) \over 2}
{1 \over c^3}  
{\partial c\over \partial \mu}
T^3 ,
\end{equation}
which vanishes in the limit of $T \to 0$. Thus, 
we recover the $d$-wave equivalent of 
Leggett's variational results~\cite{leggett-80a,leggett-80b} at $T=0$
for the saddle point and number equations in the context 
of $^3{\rm He}$.

Denoting $\xi_{\bf k}$ and $E_{\bf k}$ 
by 
$\xi$ and $E$, 
and 
$\xi_{{\bf k}+{\bf q}}$ and $E_{{\bf k}+{\bf q}}$  
by 
$\xi'$ and $E'$,
the matrix elements of $\mathbf{M}$ become
\begin{equation}
\begin{array}{l}
\nonumber
M_{11}^E(q) = {1\over\lambda_d} + \sum_{\bf k} 
\left(EE'+\xi\xi'\over 2EE'\right) 
\left((E+E') \Gamma^2({\bf k+q/2}) \over (iq_n)^2 - (E+E')^2 \right) ,  \\[5mm]
M_{11}^O(q) = - \sum_{\bf k} \left(\xi'E + \xi E'\over 2EE' \right)
\left(\Gamma^2({\bf k+q/2})\over (iq_n)^2-(E+E')^2 \right) iq_n , \\[5mm]
M_{12}(q) = 
- \sum_{\bf k} \left(\Delta_0^2 \Gamma^2({\bf k+q/2}) \over 2EE'\right)
\left((E+E') \Gamma({\bf k}) \Gamma({\bf k+q}) \over (iq_n)^2-(E+E')^2 \right) ,
\end{array}
\end{equation}
where the element 
$M_{11}(q) = M_{11}^E(q)+M_{11}^O(q)$ 
was split into an even ($E$) and an odd ($O$) part in $iq_n$, 
and $M_{21}(q) = M_{12}(q)$ and $M_{22}(q) = M_{11}(-q)$. 
Performing the analytic continuation $iq_n \to \omega + i0^+$ and 
defining $\eta(q) = |\eta (q)| \exp (i\phi (q)) 
= (\lambda(q)+i\theta(q))/\sqrt{2}$, 
the correction to the effective action assumes the form
\begin{equation}
\left(\lambda^* \,\,\, \theta^*\right)
\left(
\begin{array}{cc}
M_{11}^E + M_{12}  &  iM_{11}^O \\[3mm]
-iM_{11}^O        &  M_{11}^E - M_{12}
\end{array}
\right)
\left(
\begin{array}{c}
\lambda \\[3mm]
\theta
\end{array}
\right),
\end{equation}
where $\lambda(q)$ and $\theta(q)$ are real functions that may be identified 
with amplitude and phase fluctuations, respectively.
Since the excitation spectrum is gapped for $\mu < 0$, one can make a direct 
small ${\bf q}$ and $\omega$ expansion, resulting in  
$M_{11}^E + M_{12} = A - B\omega^2 +Q |{\bf q}|^2 +\dots$,  
$M_{11}^O = C\omega+\dots$, and 
$M_{11}^E - M_{12} = - D\omega^2 + R |{\bf q}|^2 +\dots,
$
where the coefficients are given by 
$A = \sum_{\bf k} \Gamma^4({\bf k}) \Delta_0^2 / 2E^3$, 
$B = \sum_{\bf k} 2 \xi^2 \Gamma^2({\bf k}) / 16 E^5$,
$C = \sum_{\bf k} \xi \Gamma^2({\bf k}) / 4E^3$, 
$D = \sum_{\bf k} 2 \Gamma^2({\bf k}) / 16 E^3$,
etc.


\section{Electronic Compressibility and Phase Diagram}

\begin{figure}
\begin{center}
\includegraphics[width=6.75cm]{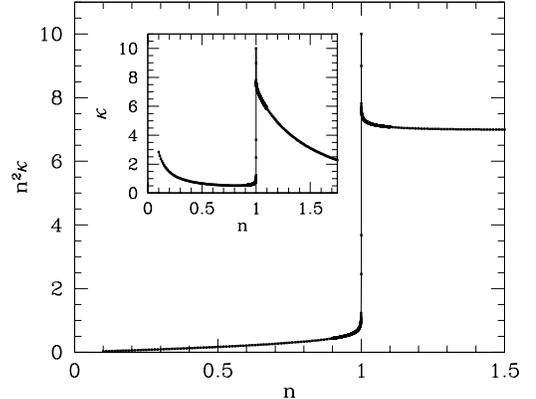}
\end{center}
\vspace{-5mm}
\caption{\small Plot of $n^2\kappa$ (in units of 
$n_{\rm max}/2\pi\epsilon_{F_{\rm max}}$)
and the electronic compressibility $\kappa$ 
(in units of $2\pi/n_{\rm max}\epsilon_{F_{\rm max}}$)
as functions of particle density 
(in units of $n_{\rm max}/2\pi$) for fixed interaction 
strength and $k_0 = k_1 = 10 k_{F_{\rm max}}$.}
\label{fig:compressibility}
\end{figure}

\begin{figure}
\begin{center}
\includegraphics[width=6.75cm]{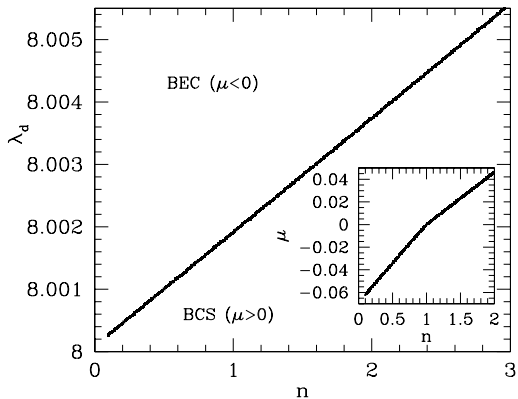}
\end{center}
\vspace{-5mm}
\caption{\small Phase diagram of particle density $n$ 
(in units of $n_{\rm max}/2\pi$) versus interaction strength $\lambda_d$ 
(in units of $g_{\rm 2D}^{-1}$, where $g_{\rm 2D}$ is the density of states
in two dimensions) for $k_0 = k_1 = 10 k_{F_{\rm max}}$.
The solid line ($\mu=0$) 
separates a gapless regime ($\mu>0$) from a fully gapped regime ($\mu<0$).
{\it Inset:} Chemical potential $\mu$ (in units of $\epsilon_{F_{\rm max}}$) 
as function of particle density $n$ across the transition region. Notice that 
$\mu$ vanishes at $n=n_c=1$.}
\label{fig:phase-diagram}
\end{figure}

We now discuss the behavior of the electronic compressibility
and phase diagram of the system as the $\mu = 0$ point is crossed.
A simple analysis of $E_{\bf k}$
indicates that, for $\mu > 0$, $E_{\bf k}$ is gapless,
while for $\mu < 0$, $E_{\bf k}$ is gapped. Simultaneously,
there is a massive rearrangement of the momentum distribution 
$n_{\bf k}$ as $\mu$ passes through $\mu_c = 0$, leading to the vanishing
of the first derivative of $\mu$ with respect to the density $n = N/L^2$
at $n = n_c$. 
An important thermodynamic quantity that depends directly on both  
$n_{\bf k}$ and $E_{\bf k}$ is the isothermal electronic 
compressibility $\kappa$, defined by
$n^2 \kappa = \left[- \partial^2 \Omega / \partial \mu^2 \right]/L^2$.
This quantity can be written as
\begin{equation}
n^2 \kappa = \alpha_0 + \alpha_{\rm fluct},
\end{equation}
where $\alpha_0 = {\partial n_0 / \partial \mu}$
and $\alpha_{\rm fluct} = {\partial n_{\rm fluct} / \partial \mu}$.
The term $\alpha_0$ at $T = 0$ can be explicitly rewritten as
$\alpha_0 = (4/L^2) \sum_{\bf k} n_{\bf k}(1-n_{\bf k})/ E_{\bf k}$,
while the term $\alpha_{\rm fluct} \propto T^3$ vanishes as $T \to 0$.
In the vicinity of $\mu = \mu_c = 0$, $\kappa$ diverges logarithmically
at $T = 0$ as
\begin{equation}
\kappa \approx 
\left[ -c_1 \ln|1 - n/n_c| + c_2 \right]
\end{equation}
in the $d$-wave case, 
suggesting the existence of a quantum phase transition (QPT). 
This singular behavior of the compressibility is shown in 
Fig.~\ref{fig:compressibility}, and the corresponding phase diagram is shown
in Fig.~\ref{fig:phase-diagram}, together with a plot of $\mu$ as 
function of $n$ (inset). The critical line $\mu = \mu_c = 0$
in the $\lambda_d \times n$ space separates a BCS-like region,
where Dirac points exist and there are gapless excitations ($\mu > \mu_c$), 
from a BEC-like region which is fully gapped ($\mu < \mu_c$).
This critical line ends at the two-body bound state threshold 
$\lambda_{d}^* = 8$ (see Fig. 3) when $n \to 0$. Thus, the QPT between the
BEC and BCS regimes requires that $\lambda_d > \lambda_{d}^*$. 
In the $s$-wave case there is no QPT,
since $\kappa$ is continuous and changes in $n_{\bf k}$ are 
always smooth.
Notice that the contribution of the collective modes to $n$ and 
to $\kappa$ at $T = 0$ vanishes identically. Thus, Gaussian fluctuation 
effects of the superconducting order parameter are not important for 
the $T=0$ electronic compressibility. 
This divergence in the second derivative of $\Omega$ 
at $T = 0$ signals a second order quantum phase transition, 
according to Ehrenfest's classification. 
However, the symmetry of the order parameter does  not
change, and a Landau symmetry classification of the phase
transition is not possible. So, if the symmetry is not changing at
the transition point, what is?


\section{The Lifshitz Transition}

To answer this question we make an immediate connection 
to the Lifshitz transition~\cite{lifshitz-60} in the context of 
ordinary metals at $T = 0$ and 
high pressure. The Lifthitz {\it transition} should not be confused with the
Lifshitz {\it point}~\cite{lifshitz-41}, 
where a {\it finite} temperature
phase transition occurs separating the high temperature disordered
phase, the spatially uniform ordered phase, and the spatially 
modulated ordered phase. 
In the conventional Lifhitz transition, the Fermi surface 
$\epsilon ({\bf k}, P) = E_F$ changes its topology as the pressure
$P$ is changed. For an isotropic pressure $P$, the deviation 
$\Delta P = P - P_c$ from the critical pressure $P_c$ is proportional to 
$\Delta\mu = \mu - \mu_c$. A typical example of the Lifshitz transition
is the disruption of a {\it neck} of the Fermi surface which leads to 
a non-analytic behavior of the number of states $N (\mu)$ 
inside the Fermi surface.
In this case, $N (\mu)$ behaves as 
$A(\mu_c) + B |\mu - \mu_c|^{3/2}$ for $\mu < \mu_c$, 
and as $A(\mu_c)$ for $\mu > \mu_c$, 
in the vicinity of $\mu_c$. 
Here, $\kappa = (3/2) B |\mu - \mu_c|^{1/2}/n_c^2$, 
where $n_c = N_c/V$.
Notice that $\kappa$ is non-analytic, but it is not singular.
The quantity that signals a phase transition in this case is not 
$\kappa$, but the thermopower $Q$, which is proportional 
to $-\partial \ln (n)/\partial \mu$,
thus leading to $Q \propto - |\Delta \mu|^{-1/2}$. 
In the conventional Lifshitz transition,
the system lowers its energy by 
$\Delta E \propto  - |\Delta \mu|^{5/2} \propto  - |\Delta P|^{5/2}$,
and the transition is said to be of second-and-half 
order~\cite{abrikosov-88}.  

For the Lifshitz transition in $d$-wave superconductors, there is already 
a non-analytic and singular behavior in $\partial n /\partial \mu$, 
and thus in $\kappa$. 
The singularity in $\kappa$ is logarithmic as indicated above, 
since the system lowers its energy by 
$\Delta E \propto  - |\mu - \mu_c|^2 \ln| \mu - \mu_c|$.
This logarithmic contribution originates from the simultaneous 
collapse of the four Dirac points at ${\bf k = 0}$, which
produces a gap in the excitation spectrum $E_{\bf k}$ 
and a massive discontinuous rearrangement
of the momentum distribution $n_{\bf k}$ in the ground state
as $\mu \to \mu_c = 0$. 
A direct topological analogy with the standard Lifshitz 
transition can be made by noticing the collapse of the 
Dirac cones at $\mu = \mu_c$ (and the disruption of a {\it neck}
for $\mu < \mu_c$) in the excitation spectrum of the system,
as shown in Fig.~\ref{fig:excitation-energy} .


\section{Superfluid Density}
\begin{figure}
\begin{center}
\includegraphics[width=6.75cm]{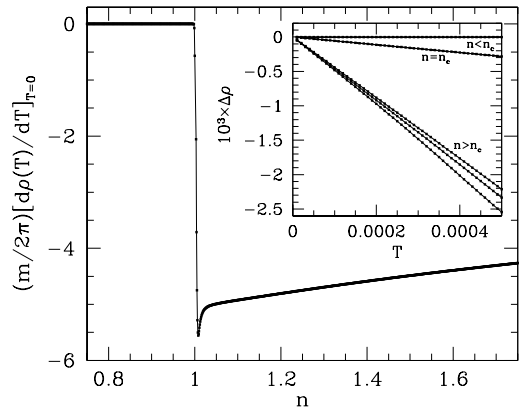}
\end{center}
\vspace{-5mm}
\caption{\small Plot of the superfluid density zero 
temperature slope (in units of 
$n_{\rm max}/2\pi\epsilon_{F_{\rm max}}$) 
as a function of $n$ (in units of $n_{\rm max}/2\pi$).
{\it Inset:} $\Delta\rho$ (in units of $n_{\rm max} / 2\pi$) 
as a function of $T$ (in units of $\epsilon_{F_{\rm max}}$) 
for various particle densities.}
\label{fig:superfluid-density}
\end{figure}

Now we analyze the behavior of the superfluid density tensor 
$\rho_{ij} (T, n)$ across the Lifshitz transition, given by 
\begin{equation}
\label{eqn:superfluid-density}
\rho_{ij} (T) = {1 \over L^2} 
\sum_{\bf k}
\left[ 2 n_{\bf k}
{\partial_i} {\partial_j} \xi_{\bf k}
-
Y_{\bf k}
{\partial_i} \xi_{\bf k} {\partial_j} \xi_{\bf k} 
\right],
\end{equation}
where $n_{\bf k}$
is the momentum distribution and 
$
Y_{\bf k} = (2T)^{-1} 
{\rm sech}^2 \left( { E_{\bf k} / 2 T} \right)
$
is the Yoshida distribution. 
Notice that $\rho_{xx} = \rho_{yy} \equiv \rho$,
while $\rho_{xy} = \rho_{yx} = 0$. Due to Galilean invariance
of our continuum model, $\rho (T = 0) = n/m$ is
well-behaved as the critical point is crossed.
However, $\partial \rho (T = 0) / \partial \mu = n^2 \kappa / m$
diverges at $n = n_c$. Using our energy and momentum scales, we define 
$\Delta\rho \equiv (m\rho (T) - n)/2\pi$. 
In Fig.~\ref{fig:superfluid-density}, we show
$\Delta\rho$ for various concentrations (inset), 
and the zero temperature slope 
$(m/2\pi) [d\rho/dT]_{T=0}$ as a function of particle 
density. Notice the discontinuity in the zero 
temperature slope as the critical point $n = n_c$ is 
approached. 

We have also calculated $\Delta\rho$ analytically 
at low temperatures in the case of very short range 
interactions ($k_0 \to \infty$). 
In the BCS limit,
$\Delta\rho = - 2 T k_1^2 / \Delta_0 (n)$,
due to the nodal structure of the $d$-wave symmetry. 
At the critical point $\mu = 0$ $(n = n_c)$,
$\Delta\rho = - \ln (2) F(\eta) T$, where
$F(\eta) = (1+\eta^2)^{-1/2}$,
with $\eta = \Delta_0 (n)/ k_1^2$.
In the BEC limit, 
$\Delta\rho = - (8/\pi) \exp(-|\mu|/T) F(\eta)$. 
This exponential behavior reflects the appearance of 
a full gap in the excitation spectrum for $n < n_c$. 

For all values of $\mu$, there is further reduction 
of $\rho (T)$ at low $T$ due to Goldstone
modes leading to  $\Delta \rho_G = - A T^3$, which obviously
does not affect the $T =0$ slope of $\Delta \rho$ through $n = n_c$.


\section{Summary} 

We have proposed the existence of a 
Lifshitz transition at $T = 0$ in clean $d$-wave superconductors,
where a topological change in momentum space 
is responsible for a non-analytic behavior of the  
electronic compressibility and of the zero temperature slope of 
the superfluid density.
We conclude by suggesting that the search for this 
transition may now be possible using the so-called ferroelectric
field effect transistor (FFET)~\cite{ahn-03},
where some control over the particle density in cuprate
superconductors may be achieved without chemical doping.
In addition, it may be possible to investigate the occurrence  
of this transition by measuring the superfluid density 
as a function of doping~\cite{lemberger-02}, 
or through a direct measurement of the electronic 
compressibility as a function of particle 
density, as was done in the study of metal-insulator 
transitions~\cite{ilani-00}.

\section*{ACKNOWLEDGMENTS} 

We thank A. J. Leggett and A. A.
Abrikosov for references and comments. We also thank
NSF (Grant No. DMR 0304380) for support.

\end{document}